\documentclass[runningheads]{llncs}
\usepackage[T1]{fontenc}

\usepackage{amssymb}
\usepackage{booktabs}
\usepackage{graphicx}
\usepackage{color}
\usepackage{comment}
\usepackage{amsmath}
\usepackage{graphicx}
\begin{document}
\title{Image Synthesis-based Late Stage Cancer Augmentation and Semi-Supervised Segmentation for MRI Rectal Cancer Staging}
\titlerunning{Image synthesis-based late stage cancer augmentation}
%
\author{Saeko Sasuga\inst{1} \and
Akira Kudo\inst{1} \and
Yoshiro Kitamura\inst{1} \and
Satoshi Iizuka\inst{2} \and
Edgar Simo-Serra\inst{3}\and
Atsushi Hamabe\inst{4}\and
Masayuki Ishii\inst{4}\and
Ichiro Takemasa\inst{4}
}
%
\authorrunning{S. Sasuga et al.}
%

\institute{Imaging technology center, Fujifilm corporation, Tokyo, Japan \and
Center for Artificial Intelligence Research, University of Tsukuba, Ibaraki, Japan \and
Department of Computer Science and Engineering, Waseda University, Tokyo, Japan \and
Department of Surgery, Surgical Oncology and Science, Sapporo Medical University, Hokkaido, Japan\\ 
\email{saeko.sasuga@fujifilm.com}
}

%
\maketitle              
\begin{abstract}
Rectal cancer is one of the most common diseases and a major cause of mortality. For deciding rectal cancer treatment plans, T-staging is important. However, evaluating the index from preoperative MRI images requires high radiologists’ skill and experience. Therefore, the aim of this study is to segment the mesorectum, rectum, and rectal cancer region so that the system can predict T-stage from segmentation results.

Generally, shortage of large and diverse dataset and high quality annotation are known to be the bottlenecks in computer aided diagnostics development. Regarding rectal cancer, advanced cancer images are very rare, and per-pixel annotation requires high radiologists’ skill and time. Therefore, it is not feasible to collect comprehensive disease patterns in a training dataset. To tackle this, we propose two kinds of approaches of image synthesis-based late stage cancer augmentation and semi-supervised learning which is designed for T-stage prediction. In the image synthesis data augmentation approach, we generated advanced cancer images from labels. The real cancer labels were deformed to resemble advanced cancer labels by artificial cancer progress simulation. Next, we introduce a T-staging loss which enables us to train segmentation models from per-image T-stage labels. The loss works to keep inclusion/invasion relationships between rectum and cancer region consistent to the ground truth T-stage.
The verification tests show that the proposed method obtains the best sensitivity (0.76) and specificity (0.80) in distinguishing between over T3 stage and underT2. In the ablation studies, our semi-supervised learning approach with the T-staging loss improved specificity by 0.13. Adding the image synthesis-based data augmentation improved the DICE score of invasion cancer area by 0.08 from baseline. We expect that this rectal cancer staging AI can help doctors to diagnose cancer staging accurately.

\keywords{Rectal cancer \and Segmentation \and T stage discrimination \and Semi-supervised learning  \and Image synthesis.}
\end{abstract}
%
%
\section{Introduction}
T-staging is an index for assessing the spread of primary tumors into nearby tissues. Regarding rectal cancer which is a common disease and a major cause of mortality. Magnetic resonance imaging (MRI) is commonly used for the assessment of T2/T3/T4 staging. In the clinical research~\cite{cho2014prognostic,taylor2014preoperativeMRF} , it has become clear that invasion depth from rectum and margin between cancer and mesorectum are directly related to the prognosis of rectal cancer, and those indexes are also important for deciding rectal cancer treatment plan. However, evaluating T-staging from preoperative MRI images requires high radiologists’ skill and experience. In~\cite{rafaelsen2008transrectal}, the T3 staging sensitivity and specificity for experienced gastrointestinal radiologists is 96\% and 74\%, while for a general radiologist is 75\% and 46\%, respectively. Since those numbers are not sufficiently high, computer aided diagnosis (CAD) systems for assisting T-staging were proposed in the previous studies~\cite{kim2019rectal}.

This paper deals with two difficulties in building such CAD systems. First one is that preparing per-pixel annotation requires high radiologists’ skill and time cost. Second one is that advanced cancer cases are very rare. Due to these difficulties, it is almost impossible to collect large scale high quality training dataset. To tackle these problems, we propose semi-supervised learning and image synthesis-based late stage cancer augmentation. Here we pick up related works, then summarize contributions of this work.
 
\subsubsection{Cancer Segmentation and T-staging} 
Since the U-net~\cite{ronneberger2015unet} was proposed, segmentation tasks have achieved remarkable results. Regarding tumor segmentation tasks from MR images, it is reported that segmentation accuracy reached human expert level~\cite{feng2020brain,hodneland2021automated,huang20203DRectum}. However, not so many researches have been done on automatic T-staging. One research~\cite{dolz2018multiregion} tried to segment tumor, inner wall, and outer wall to visualize cancer invasion, but it did not evaluate T-staging accuracy. Kim et al.~\cite{kim2019rectal} proposed a method for classifying T-stage from preoperative rectal cancer MRI, but their method was not able to visualize the basis of the decision. To the best of our knowledge, there was no previous study that simultaneously evaluate both cancer segmentation and T-staging accuracy.

\subsubsection{Data augmentation}
Supervised learning methods rely on a large and diverse training dataset to achieve high performance. Data augmentation is commonly known as an effective way to compensate limited amount of training data~\cite{shorten2019surveyDA}.Recently, generation-based approach using generative adversarial networks (GANs)~\cite{goodfellow2014generative} have been proposed. One such method is label-to-image translation. Abhishek et al.~\cite{abhishek2019mask2lesion} used label-to-image translation method to generate skin cancer 2D images for data augmentation and showed improvement in segmentation accuracy. However, this method has limitations that the generated data follow training data distribution. It is difficult to create images that are rarely included in training data. \\
We summarize the contributions of this work as follows:

\begin{enumerate}
\item This is an early work which develops and evaluates a CAD system that can do both rectal cancer segmentation and T-staging with a clinical dataset.
\item We propose a semi-supervised learning using a novel T-staging loss. The T-staging loss enables us to train a segmentation model with not only per-pixel labels but per-image T-stage labels.
\item We propose a label-to-image translation based severe cancer augmentation. This approach can generate  rare data that is out of training dataset distribution by deforming cancer label into the shape of a progressive cancer.
\end{enumerate}

\section{Methodology}
\begin{figure}[t]
\centering
\includegraphics[width=0.95\linewidth]{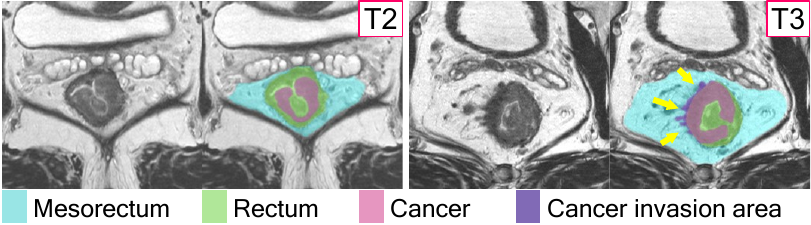}
\caption{T-stage classification algorithm using multi-label relationship. Yellow arrows point cancer
area extends beyond the rectum area and this case is staged with T3.} \label{fig:F1}
\end{figure}

We model the rectal cancer T-staging problem as a multi-class multi-label segmentation of mesorectum, rectum, and rectal cancer as shown in Figure~\ref{fig:F1}. The algorithm predicts T-stage following to the segmentation results. When the cancer area is not in contact with the contour of the rectum area, the case is classified as under T2 stage. On the other hand, when the cancer area crosses over the rectum area, the case is classified as over T3 stage. This rule exactly follows the T-staging rules of tumor invasion into the area of the rectum. In this section, we explain a semi-supervised approach with a novel loss function named T-staging loss in Section 2.1. Then, we explain image synthesis-based late stage cancer augmentation in Section 2.2.

\subsection{Semi Supervised Learning with T-staging Loss}
\begin{figure}[t]
\centering
\includegraphics[width=\linewidth]{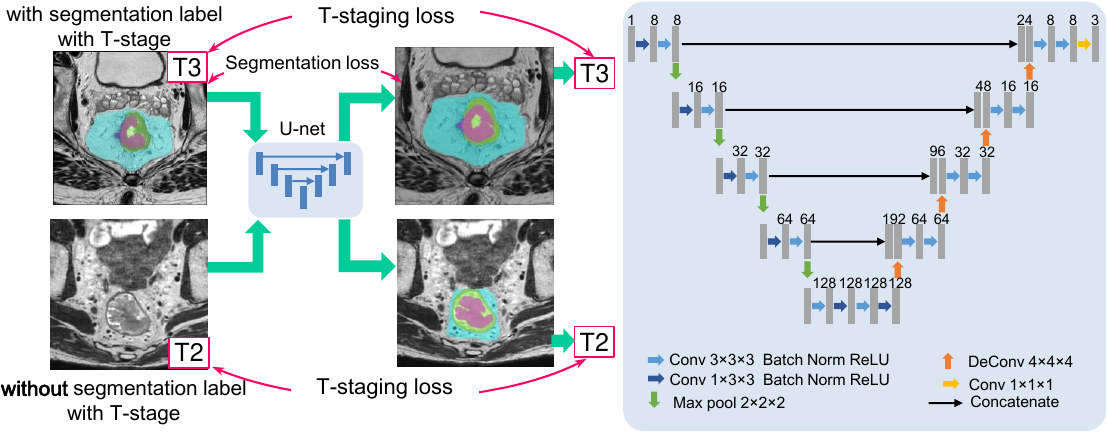}
\caption{ Frameworks of our semi-supervised learning and the architecture of segmentation
network. In the training phase, images with and without segmentation labels are inputted to
the U-net. Training data with only T-stage ground truth are used for calculating T-staging
loss which works as to maximize correspondence between ground truth and prediction T-stage.} \label{fig:F2}
\end{figure}

The proposed semi supervised learning architecture is shown in Figure~\ref{fig:F2}. We use a 3D variant of U-net that we train to perform a segmentation task. The last three channels of the network are the probabilities of mesorectum, rectum and rectal cancer areas. T-staging results are calculated based on the binarized segmentation results. In the training phase, 3D MR images with ground-truth segmentation labels or just only ground truth T-stage are inputted to the network. The loss function consists of a segmentation loss and the proposed T-staging loss;
$Loss_{SEG} + \lambda \times Loss_{STG}$,
where $\lambda$ is a parameter used to balance  the two terms. We used a standard Dice loss~\cite{milletari2016vnet} for $Loss_{SEG}$.  The T-staging loss which for accurate staging purposes is defined as follows:

\begin{equation}
\label{eq:a}
	Loss_{STG} = -\mu \log{\frac{{p_{STG}}^{g_{STG}}}{(1-p_{STG})^{(g_{STG}-1)}}}+ \frac{\sum_{i=1}^Np_{invT2i}+\alpha}{\sum_{i=1}^Np_{invT2i}+\sum_{i=1}^Np_{invT3i}+\alpha}.
\end{equation}

Where,
\begin{equation}
\label{eq:b}
	p_{STG} =\max(p_{cancer,i}\times (1-p_{rectum,i})).
\end{equation}
\begin{equation}
\label{eq:c}
g_{STG}=
\begin{cases}
1 & \text{if ground truth T stage is over T3} \\
0 & \text{otherwise}
\end{cases}.
\end{equation}
\begin{equation}
\label{eq:d}
	p_{invT2} = p_{cancer}\times(1-p_{rectum})\times(1-g_{staging}).
\end{equation}
\begin{equation}
\label{eq:e}
	p_{invT3} = p_{cancer}\times(1-p_{rectum})\times(1+g_{staging}).
\end{equation}

\noindent N is the number of voxels. $p_{cancer}$ and $p_{rectum}$ represent the probability maps of the rectal cancer and rectum, respectively. $p_{staging}$ indicates the probability of the predicted staging. It takes a high value when there is any voxel simultaneously having low rectum probability and high cancer probability. $p_{invT2}$ and $p_{invT3}$ are probabilities of cancer invasion of each voxel in T2 case and T3 case. This term works to reduce the cancer area outside of the rectum for T2 cases. On the other hand, it works to increase the cancer area outside of the rectum for T3 cases. In contrast to $Loss_{SEG}$ is used only for the images with ground-truth segmentation labels. $Loss_{STG}$ can be used for all cases having only ground truth T-staging. So, it works as semi-supervised setting in segmentation.

\subsection{Generating Advanced Cancer MRI Image from Labels}
\begin{figure}[t]
\centering
\includegraphics[width=\linewidth]{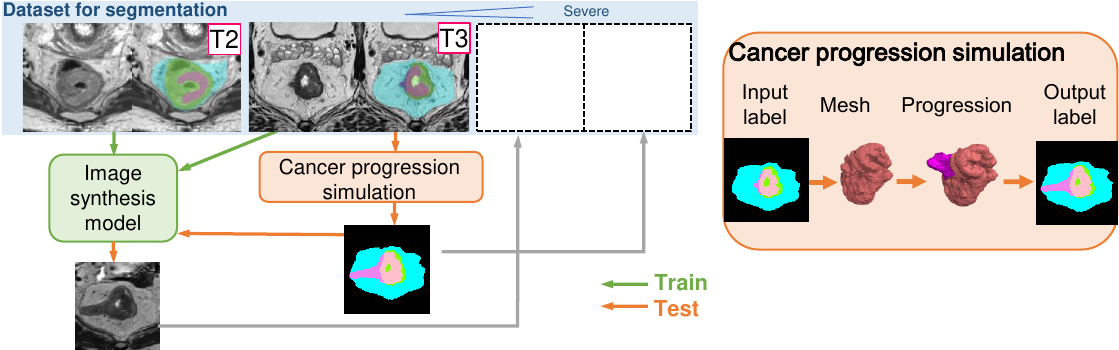}
\caption{Overview of severe cancer data syntheses framework. The image synthesis model is
trained with real images. On the test phase, cancer labels are modified by simulation. Following the generated labels, image synthesis model generate severe cancer images.} \label{fig:F3}
\end{figure}

Figure~\ref{fig:F3} is an overview of advanced cancer image synthesis system. In general, GAN can model the underlying distribution of training data. To the contrary, our goal is to generate out-of-distribution data (e.g., we have relatively many early stage cancer, but we want more advanced staged cancer images). Therefore, we first train a label-conditional-GAN model. Then we generate images for data augmentation by inputting various deformed label images. An important point is that the deformation is done by external knowledge which simulate cancer development. 

\subsubsection{Semantic image synthesis}
SPADE~\cite{park2019semantic} is known as one of the best architectures for the semantic image synthesis task. However, it is rarely applied on 3D images due to their large computational cost. This is an early work applying SPADE for 3D image synthesis. Overall, we adopt and extended the basic framework of SPADE from 2D to 3D. As shown in Figure~\ref{fig:F4}, we simply introduce a SPADE-3D block in each scale of the generator instead of residual-SPADE block which was proposed in the original paper to save computational cost.
 
We found that the more the kinds of input label classes, the more generated images give realistic looking and anatomically consistent. For that reason, in addition to cancer, we also used rectum, mesorectum, bladder, prostate and pelvis classes as the input to the generator. We monitored the training until the generator gives realistically looking images from the given test label images.

\subsubsection{Cancer label deformation}
\begin{figure}[t]
  \centering
    \begin{tabular}{c}
       \begin{minipage}{0.7\hsize}
          \includegraphics[width=1.0\linewidth]
                          {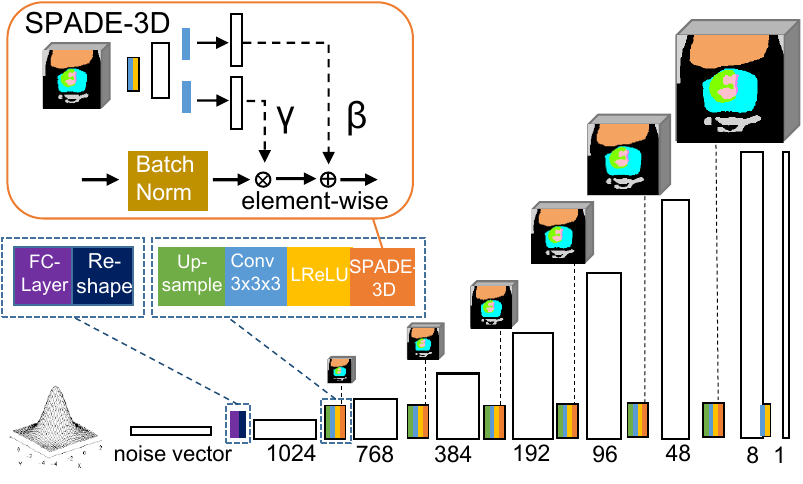}
       \end{minipage}
      \begin{minipage}{0.3\hsize}
        \centering
          \includegraphics[ width=1.0\linewidth]
                          {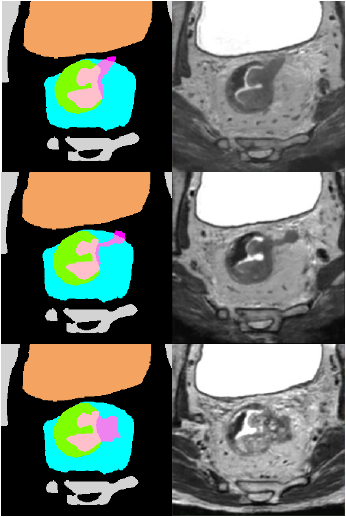}
      \end{minipage} \\ 
 \begin{minipage}{.69\hsize}
        \caption{Overview of the 3D generator architecture. Each normalization layer uses 3D label image to modulate the layer activations.
To save computation cost, we removed residual blocks in
the original SPADE~\cite{park2019semantic} and introduced a single SPADE-3D block
per scale which had enough capacity to generate realistic images.}
\label{fig:F4}
    \end{minipage}%
    \hfill%
	 \begin{minipage}{.29\hsize}
        \caption{Samples of the input labels and generated images. Different images are generated based on the deformed cancer labels.}
\label{fig:F5}
    \end{minipage}%
    \end{tabular}
\end{figure}

Advanced stage cancer labels are generated as follows. First, a cancer label was transformed to a 3D mesh model consisting of vertices and faces. 
Next, the furthest invasion point, infiltration direction vector, and non-deformable part were calculated based on cancer morphology~\cite{RectumCancerStaging}. Then, the 3D mesh model was transformed by constraint shape optimization~\cite{botsch2004intuitive}. Finally, the mesh model was voxelized again. These processes were applied several times for each image, randomly changing a vertex to move and shifting distance. We also mimicked the shape of the cancer infiltrate into vessels or lymph nodes by randomly embedded the tubular or spherical objects to the cancer model.
 The MR images  corresponding to these labels were variously generated by the GAN as shown in Figure~\ref{fig:F5} and supplementary movies.

\section{Experiments and Results}

\subsection{Dataset}
 We curated four types of datasets A-D. The dataset A-C were real MR images acquired from rectal cancer patients. The dataset D was generated by a GAN model which was trained on dataset A. The dataset B was separated from A for a fair comparison, meaning that the dataset D was not generated from testing dataset B. The dataset A, B, and D had per-pixel ground truth labels, and dataset C had only T-stage label. The number of samples in each dataset are provided in Table~\ref{tb:tab1}.

MR T2-weighted images were acquired using a 3.0 T (N=89) or 1.5 T (N=106) MR scanner. Ground truth segmentation labels were annotated by two surgeons based on the pathological specimens removed by surgeries. Ground truth T-stage labels were labeled based on the pathological diagnosis.

\begin{table}[t]
\caption{ The dataset specification and usage in this study.}
\label{tb:tab1}
\centering
\setlength{\tabcolsep}{3pt}
\vspace{-1.2mm}
\begin{tabular}{llccccc}
\toprule
\multicolumn{1}{}{}&\multicolumn{1}{l}{Patient Num.}  & \multicolumn{2}{l}{Data type}     & \multicolumn{2}{l}{Usage}     \\ 
 &(T2/T3)&Image type&Ground truth& Segmentation&Image Synthesis\\
\hline
A&	69 (22/47)&	Real&	Label&	Train&	Train\\
B&	66 (26/40)&	Real&	Label&	Train/Evaluation&	Unused\\
C&	60 (29/31)&	Real&	T-Stage&	Train/Evaluation&	Unused\\
D&	78 (0/78)&	Generated&	Label&	Train&	-\\

\bottomrule
\end{tabular}
\end{table}

\subsection{Implementation Details and Evaluation Metrics}
During the training, the Adam optimizer with a learning rate of 0.003 was used. The parameters in the loss function are as follows; $\lambda $=0.1,  $\mu $=0.1,$\alpha $=500. Mini batch consisted of 4 cases with segmentation labels and 2 cases with only T-stage.  Experiments were conducted up to 80000 iterations.

For the evaluation, five-fold cross validation was conducted on dataset B and C. Those evaluation datasets were randomly divided into five subgroups. Four subgroups out of the five and dataset A, and D were used for training the segmentation network. Note that one subgroup in the four subgroups for training was used for validation. We chose the best model according to the DICE score of the validation subgroup, then applied the model to the remaining subgroup. We used two indexes of Dice similarity coefficient in volume~\cite{zou2004statistical}, and T-staging sensitivity/specificity for evaluation. We calculated T-staging sensitivity as a correctly predicted T3 number divided by the ground truth T3 number. The specificity is a correctly predicted T2 number divided by the ground truth T2 number.

\subsection{Results}
Ablation studies were conducted to evaluate the efficacies of the proposed methods. We compared three experimental settings shown in Table~\ref{tb:tab2}. Each setting corresponds to the baseline (the DICE loss~\cite{milletari2016vnet}), baseline plus semi-supervised approach, and baseline plus semi-supervised approach and late stage data augmentation.

\begin{figure}[t]
  \centering
    \begin{tabular}{c}
       \begin{minipage}{0.65\hsize}
          \includegraphics[width=1.0\linewidth]
                          {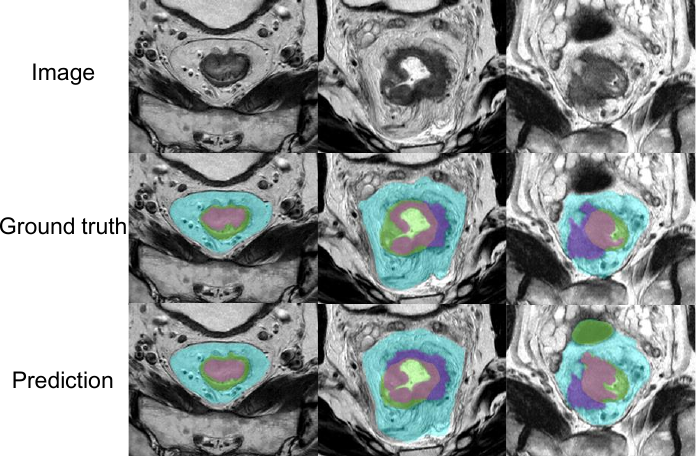}
       \end{minipage}
      \begin{minipage}{0.32\hsize}
        \centering
          \includegraphics[ width=1.0\linewidth]
                          {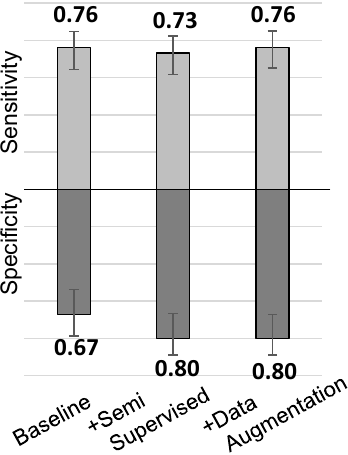}
      \end{minipage} \\ 
 \begin{minipage}{.61\hsize}
        \caption{Example segmentation results.
The left column shows T2 cases, the center column is T3 case and 
the right column is a severe cancer case.}
\label{fig:F6}
    \end{minipage}%
    \hfill%
	 \begin{minipage}{.32\hsize}
        \caption{Comparison of T-staging sensitivity and specificityin ablation study.}
\label{fig:F7}
    \end{minipage}%
    \end{tabular}
\end{figure}

We compared the three settings with the Dice coefficients of mesorectum, rectum, cancer and cancer invasion area. Table~\ref{tb:tab3} shows the results of dice coefficient. Figure~\ref{fig:F6} shows several segmentation results. Both of the semi-supervised and the data augmentation increased the DICE score of cancer area. Especially, they were effective in cancer invasion area. The segmentation results of mesorectum and rectum were not affected by the proposed approach, as is expected. Figure~\ref{fig:F7} shows results of T-staging accuracy. The proposed method obtains the best sensitivity (0.76) and specificity (0.80), and improved specificity by a margin of 0.13.

\begin{table}[t]
\caption{The condition of ablation studies.}
\label{tb:tab2}
\centering
\setlength{\tabcolsep}{6pt}
\vspace{-1.2mm}
\begin{tabular}{llll}
\toprule
 &Loss function& Dataset\\
\hline
Baseline (DICE loss)& Soft dice& A,B\\
+Semi supervised& Soft dice+T-staging loss& A,B,C\\
+Data augmentation& Soft dice+T-staging loss& A,B,C,D\\
\bottomrule
\end{tabular}
\end{table}

\begin{table}[t]
\caption{Comparison of DICE score in ablation study.}
\label{tb:tab3}
\centering
\setlength{\tabcolsep}{6pt}
\vspace{-1.2mm}
\begin{tabular}{lccccccc}
\toprule
 & Mesorectum  & Rectum & Cancer& Invasion area \\
\hline
Baseline (DICE loss)& 0.907&0.908&0.642&0.415 [0.330-0.501]\\
+Semi supervised&	\bf{0.910}&	0.910&	0.669&	0.449 [0.380-0.519]\\
+Data augmentation	&0.903&	\bf{0.911}&	\bf{0.680}&	\bf{0.495} [0.413-0.577]\\
\bottomrule
\end{tabular}
\end{table}


\section{Discussion}
We assumed that the number of available per-pixel training data is small. This often happens at clinical practice. In fact, since rectal cancer is ambiguous on MR images, the surgeons annotated labels with reference to pathological specimens to make reliable ground truth in this research. The T-staging accuracy of the proposed methods was not high enough compared to pathologically proved ground truth. One reason is that T-staging is a hard task for a radiologist too. Another one is that the proposed data augmentation handles only rare shape of the cancer. The future work is generating not only rare shape but rare texture.
This study was conducted on data acquired from one site. We are going to validate the algorithm with multi-site data and severe real cancer data.

\section{Conclusions}
In this paper, we proposed a novel semi-supervised learning method designed for cancer T-staging, and image synthesis-based data augmentation to generate advanced cancer images. Ablation studies showed that the methods improved rectal cancer segmentation and cancer T-staging accuracy. We expect that this cancer segmentation would help doctors diagnose T-staging in clinical practice.

\newpage

%
%
%
%
\bibliographystyle{splncs04}
\bibliography{mybibliography}

\end{document}


%

\begin{figure}[H]
\centering
\includegraphics[width=0.98\linewidth]{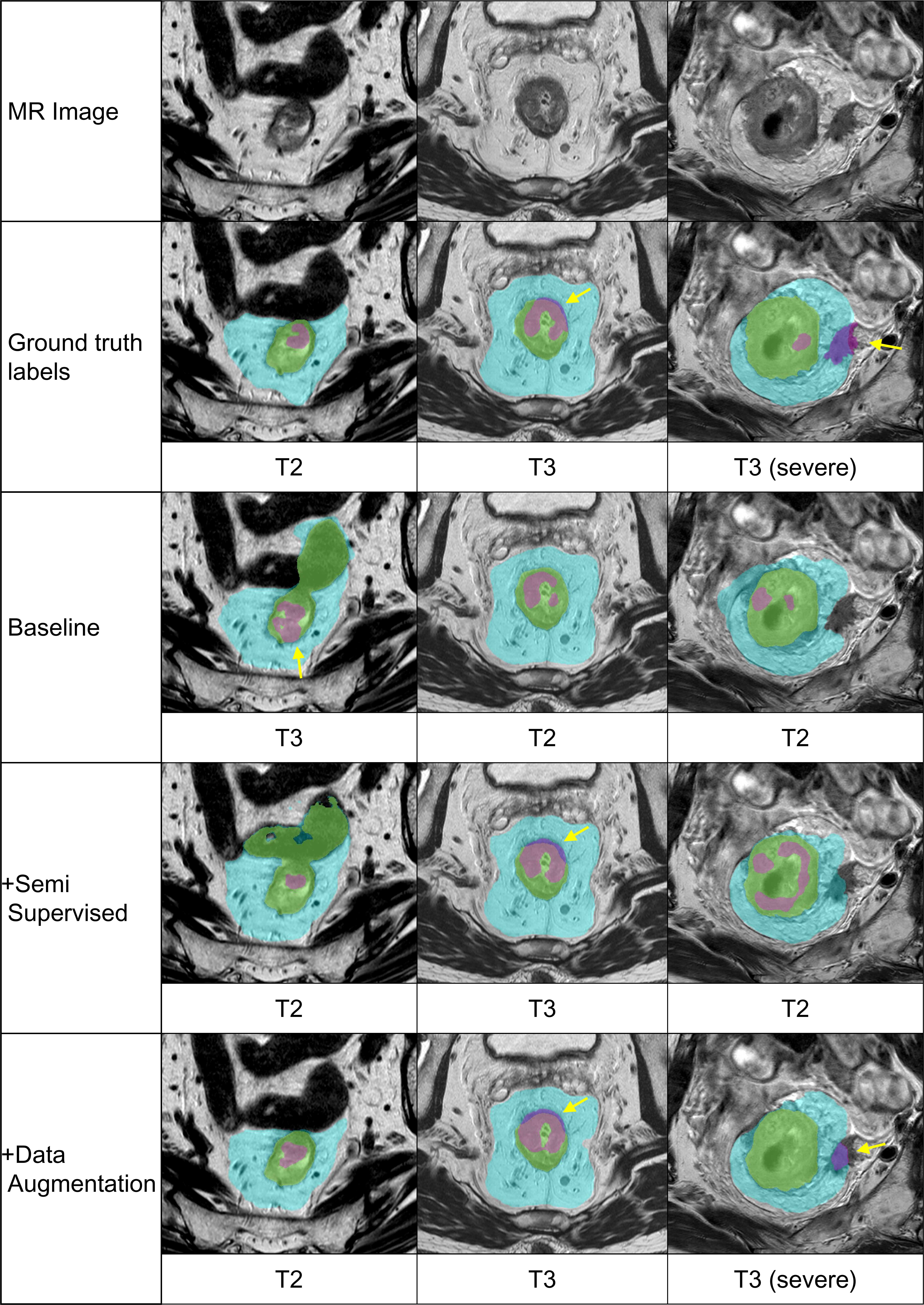}
\caption{ Examples of segmentation results and predicted T-stage.
The left and center columns show examples in which semi-supervised learning was effective in improving T-staging accuracy.
 The right column is the case in which our data augmentation method contributed to the severe cancer segmentation accuracy.
Yellow arrows point to the cancer invasion area.
} \label{fig:F1}
\end{figure}

\newpage

\begin{table}[t]

\caption{ Detailed supplementation of loss function term and evaluation metrics.}
\vspace{-1mm}
\label{tb:tab1}
\setlength{\tabcolsep}{5pt}
\centering
\begin{tabularx}{0.98\textwidth}{lxX}
\toprule
Metric& $\text{Definition}$ &\\
\hline

$Loss_{SEG}$
& {\displaystyle   1-\frac{2\sum_{i=1}^Np_{i}g_{i}}{\sum_{i=1}^Np_{i}+\sum_{i=1}^Ng_{i}}.}%
& ${N}$: the number of voxels

$ p $: the probability maps from the network

$ g$: the ground-truth label\\
DICE score
& {\displaystyle    \frac{2|P \cap G|} {|P|+|G|}.}%
& $P$: the predicted results

$G$: the ground-truth label \\

Sensitivity
& {\displaystyle  \frac{|{P_{T3}} \cap{G_{T3}}|} {|G_{T3}|}.}%
&  $P_{T3}$: the case predicted T-stage as over T3

$G_{T3}$: the case labeled T-stage as over T3\\

Specificity
& {\displaystyle \frac{|{P_{T2}} \cap{G_{T2}}|} {|G_{T2}|}.}%
&  $P_{T2}$: the case  predicted as under T2

$G_{T2}$: the case labeled as under T2\\
\bottomrule
\end{tabularx}
\end{table}

\begin{table}[h]

\caption{ Implementation and experimental configuration details.}
\vspace{-1mm}
\label{tb:tab2}
\centering
\setlength{\tabcolsep}{4pt}
\begin{tabularx}{0.98\textwidth}{ll}
\toprule
Software framework &Tensorflow19.04\\
Preprocess&  $\bullet $Rescale voxel size to 0.5 $\text{mm}^3$ isotropic. \\
		&$\bullet $Crop around the cancer area \\
		& \; to 192 × 192 × 112 voxel size in trainig phase /\\
		& \; to 256 × 256 × 128 voxel size in evaluation phase.\\
		 &$\bullet $Normalize value from 0 to1 by percentile.\\
				
Hardware&	NVIDIA DGX-2 machine \\
			&using the NVIDIA V100 GPU with 64 gb memory\\

The average training runtime&Five days\\

\bottomrule
\end{tabularx}
\end{table}

